\documentclass[prl,twocolumn,showpacs,aps]{revtex4}

\usepackage{graphicx}
\usepackage{dcolumn,float}
\usepackage{amsmath,amssymb}

\newcommand{\nwc}{\newcommand}

\nwc{\ket}[1]{|#1\rangle}
\nwc{\bra}[1]{\langle#1|}
\nwc{\scal}[2]{\bra{#1}#2\rangle}
\nwc{\be}{\begin{equation}}
\nwc{\ee}{\end{equation}}
\nwc{\bea}{\begin{eqnarray}}
\nwc{\eea}{\end{eqnarray}}
\nwc{\bb}{\boldsymbol{\beta}}
\nwc{\ba}{\boldsymbol{\alpha}}
\nwc{\cA}{{\textsf{A}}}
\nwc{\cD}{{\textsf{D}}}
\nwc{\cQ}{{\textsf{Q}}}
\nwc{\cR}{{\textsf{R}}}
\nwc{\cS}{{\textsf{S}}}
\nwc{\cT}{{\textsf{T}}}
\nwc{\cV}{{\textsf{V}}}
\nwc{\Sv}{{\mathbf{S}}}
\nwc{\Hv}{{\mathbf{H}}}

\begin{document}

\title{Continuum in the Excitation Spectrum of the $S=1$ Compound CsNiCl$_3$ }

\date{\today}

\author{E.~Ercolessi, G.~Morandi and M.~Roncaglia}

\affiliation{Physics Department, University of Bologna,
INFM and INFN, V. B.~Pichat 6/2, I-40127, Bologna, Italy.}

\begin{abstract}
Recent neutron scattering experiments on CsNiCl$_3$ reveal some features
which are not well described by the nonlinear $\sigma$ model nor by numerical
simulations on isolated S=1 spin chains.
In particular, in real systems the intensity of the continuum of multiparticle
excitations, at $T=6\; \mathrm{K}$, is about 5 times greater than predicted.
Also the gap is slightly higher and the correlation length is smaller.
We propose a theoretical scenario where the interchain interaction is
approximated by a staggered magnetic field, yielding to a correct
prediction of the observed quantities.

\end{abstract}


\maketitle


About twenty years after the famous ``Haldane conjecture''\cite{Hal83}, quantum spin chains
with Heisenberg antiferromagnetic interactions
are still attracting much experimental and theoretical interest.
Half-integer spin chains have no spin gap, and hence are critical at
$T=0$ with algebraic decay of correlations.
Integer spin chains have a gap in the excitation spectrum and the ground
state is characterized by a finite correlation length.

For S=1 chains, the excitation spectrum is dominated by a well-defined magnon
triplet carrying spin 1. These excitations have been observed in several
neutron scattering experiments and in the case of NENP, they constitute
a large part of the total spectrum.

However, recent experimental measurements on CsNiCl$_3$ reveal a significant
spectral weight in the incoherent multiparticle continuum above the coherent
magnon peak.
At $6\;\mathrm{K}$ the integrated intensity of the continuum
around the antiferromagnetic point is about 9(2)\% of the
total spectral weight\cite{Sperim01}.
This result is considerably larger than the
1-3\% weight predicted by numerical simulations\cite{numeric} and the O(3) non-linear
sigma model for the 3-particle continuum\cite{Essler}.
The effects of the coupling between chains has been considered in the framework of RPA,
but no significative increase of the continuum has been found\cite{Essler}.
On the other hand, Majorana fermion theory yields a prominent
three-particle-scattering continuum\cite{Essler}, but this
would require strong biquadratic exchange interactions of the form
$(\Sv_i\cdot\Sv_{i+1})^2$ which are not in accordance with the measurements on
CsNiCl$_3$\cite{Sperim01}.

Furthermore, some measured quantities at $T=6\; \mathrm{K}$ show appreciable deviations
from analytical and numerical predictions at finite temperature
on isolated chains\cite{Joli,Kim}, i.e.
the measured gap $\Delta$ is higher and the correlation length $\xi$ is shorter than the corresponding
theoretical values, breaking also the single mode approximation (SMA) correspondence $\Delta\xi=c$,
where $c$ is the spin-wave velocity.

The model Hamiltonian proposed for CsNiCl$_3$ is
\begin{equation}
H=J \sum_{i}^{\rm chain} \Sv_{i}\cdot\Sv_{i+1}
+J' \sum_{\langle ij \rangle}^{\rm plane}\Sv_{i}\cdot\Sv_{j}
+D\sum_{i} (S^z_i)^2\ ,
\label{hamil}
\end{equation}
where the coupling $J=2.28\;\mathrm{meV}$ along the $c$-axis
is much stronger than the exchange interaction in the basal plane
$J'=0.044\;\mathrm{meV}$\cite{Sperim}.
The single-ion anisotropy is estimated to be $D\approx 4\;\mu$eV, small enough
that CsNiCl$_3$ may be considered purely isotropic, hence we put $D=0$ from now on.
Below $T_{\rm N}=4.85$\;K the system undergoes 3D long-range ordering,
caused by the interchain interaction.
For $T>T_{\rm N}$ the system is regarded as one-dimensional and commonly considered to be a good
realization of a single isotropic spin chain, although there is a sizable dispersion
perpendicular to the chain direction due to the interchain
coupling.

Here comes our main assumption:

\noindent
\textit{In some range of intermediate temperatures above the
ordering one, $T_N<T<T^*$, the interchain coupling induces a (short-range) three-dimensional
antiferromagnetic order which varies on a length scale $\xi_{3D}$ which is greater
than the correlation length $\xi_{1D}$ of the single chain}.

In other words, a little above $T_N$, even if the total staggered magnetization is zero,
three-dimensional effects are still important and the system is arranged in large domains
with non-zero magnetization. Thus, in a mean-field picture, a single chain experiences a
staggered magnetic field generated by the neighboring chains.
No matter if the field is not constant
along the whole chain, because it varies so slowly that it may be
considered constant, since the 1D chains are gapped and short-ranged.

Consistently, we suppose the existence of a temperature $T^*$ above which the antiferromagnetic
order is so reduced that the 1D system may be viewed as isolated.
To support this hypothesis we mention that in the real compound CsNiCl$_3$, some experimentalists\cite{Sperim01}
found that for $T\gtrsim 12$\;K the measurements turn out to be
in accordance with the theoretical predictions on isolated chains.

In the mean-field approximation the interchain interaction is treated as a staggered
magnetic field that takes into account the effect of the neighboring chains on the
single 1D system
\begin{equation}
J' \sum_{\langle ij \rangle}^{\rm plane}\Sv_{i}\cdot\Sv_{j}\quad \Longrightarrow \quad
H_s \sum_i (-)^i S_{i}^z\ ,
\end{equation}
where the staggered field (we choose it along the $z$-axis) is evaluated
self-consistently by means of the fixed-point equation\cite{ST90}
\begin{equation}
H_s=z_c J'\; m_s(H_s)\ ,
\label{fp}
\end{equation}
with $z_c$ the coordination number in the basal plane and
$m_s(H_s)$ is the magnetization curve of a single chain in presence
of the external staggered field $H_s$.
In our calculation we fix the energy scale $J=1$, the interchain coupling $J'=0.02$
as close as possible to the experimental value for CsNiCl$_3$ and we put $z_c=3$ because
the lattice in the basal plane is triangular.

So, the Hamiltonian we are going to study is
\begin{equation}
\mathcal{H}=\sum\limits_{i}\left[J\;\Sv_{i}\cdot\Sv_{i+1}\; +\; (-1)^{i}\;H_s\; S_{i}^z\right]\ .
\label{ham}
\end{equation}

Following the Haldane mapping \cite{libri1}, we represent the local spin operators as
$\mathbf{S}_{i}\approx S(-1)^{i}\mathbf{n}_{i}+\mathbf{l}_{i}$, with $\mathbf{n}_{i}^{2}=1$,
where $\mathbf{n}_{i}$ represents the slowly-varying local staggered magnetization and
$\mathbf{l}_i$ is the local generator of angular momentum.
In this framework, the Zeeman term of Eq.(\ref{ham}) becomes essentially a linear shift in the
$\mathbf{n}$-field along the $z$-direction,
$(-1)^i H_s S_i^z\approx S\; H_s n_i^z$.

Going then to the continuum limit and integrating out the fluctuation field
$\mathbf{l}$ we obtain the O(3) NL$\sigma$M plus a linear term\cite{EMPR_nov00}
\begin{equation}
\mathcal{L}=\frac{1}{2gc}\left( c^{2}|\partial_{x}\mathbf{n}
|^{2}+|\partial_{\tau}\mathbf{n}|^{2}\right) -S\; H_s n^z -i\lambda\left(\mathbf{n}^{2}-1\right)\ ,
\end{equation}
where $g=2/S$ and $c=2JS$. A Lagrange multiplier $\lambda(x,\tau)$ has been introduced to implement
the local constraint $\mathbf{n}^{2}(x,\tau)=1$.

The experiments were performed at temperatures small enough if compared with
the exchange interaction $J$ along the chain, so we mainly consider the zero-temperature
case arguing that a calculation at finite temperature\cite{Joli} will give simply a
small correction in the calculated quantities.

For a constant staggered field, the associated saddle point will
correspond to a constant value of $\lambda$ and the self-consistency
equation at $T=0$ is
\begin{equation}
\frac{3g}{2\pi}\ln\left\{\Lambda\xi+\sqrt{1+(\Lambda\xi)^{2}}\right\}=
1-\left(\frac{S g}{c}\right)^{2}H_s^2\xi^{4}\ ,
\label{self}
\end{equation}
where $\xi$ is the correlation length and $\Lambda$ is the momentum cutoff of the theory.
In the passage to the continuum limit we have lost the correct renormalization
of the parameters.
In order to restore them, we fix the zero-field values of $\Delta_{0}=0.41048\;J$, $\xi=6.03$
(which correspond to $c=\xi\Delta_{0}=2.48\;J$), known from the DMRG studies.
The NL$\sigma$M coupling has been left to its analytical value $g=2$.
As the field is varied, the cutoff $\Lambda$ is held fixed to its zero-field value
$\Lambda=0.2072$ and we use Eq.(\ref{self})
to determine the value of $\xi=\xi(H_s)$.

The magnetization per site is given by
\begin{equation}
m_{s}(H_s)=S\left\langle n^z \right\rangle =\frac{gS^{2}}{c} \; \xi^{2}(H_s)\;H_s  \ .
\end{equation}
Putting it back in Eq.(\ref{fp}), we find a stable solution for $H_s=0.027$, corresponding to
$m_s=0.45$, $\xi=4.54$ and $\Delta_T=c \xi^{-1}=0.545$.
It is remarkable that even for a small staggered field the response of the
system is
strong, generating an appreciable magnetization, meanwhile the gap renormalizes
upward and the correlation length is getting shorter.

The staggered field breaks explicitly the symmetry and the quasi-particle propagator
splits in one longitudinal channel (i.e. in the direction of the field) and two
transverse channels.
The transverse propagator is given by
\begin{equation}
G_T(q,\omega)=\frac{S^2 g c}{\omega^{2}+c^{2}q^{2}+\Delta_T^2}\ ,
\end{equation}
and has simple poles at $\omega=\pm\varepsilon(q)$ with
\begin{equation}
\varepsilon(q)=\sqrt{c^{2}q^{2}+\Delta_T^2}\ ,
\end{equation}
where $q=0$ corresponds to the antiferromagnetic point.
In the transverse channel the theory is purely bosonic and the spectral weight is fully
exhausted by this magnon excitation. In this case, the dynamical structure factor
$S_T(q,\omega)\equiv\operatorname{Im}G_T(q,\omega)/\pi$ is simply given by
\begin{equation}
S_T(q,\omega)=\frac{gcS^{2}}{2\varepsilon(q)}\left\{\delta(\omega-\varepsilon(q))-
\delta(\omega+\varepsilon(q))\right\}
\end{equation}

On the other hand, the longitudinal propagator must be evaluated more carefully
considering its connected part and the calculation\cite{EMPR_dic00} gives
\begin{equation}
G_L(q,\omega)= G_T(q,\omega)\;\frac{3\widetilde{\Gamma}(q,\omega)}
{3\widetilde{\Gamma}(q,\omega) + 2 m_s^{2} G_T(q,\omega)}\ ,
\end{equation}
\pagebreak
where $\widetilde{\Gamma}(q,\omega)$ is the Fourier transform of the product of
two propagators
\begin{equation}
\Gamma(\mathbf{x-x'})= \frac{1}{S^4}G_{T}(\mathbf{x-x'}) G_{T}(\mathbf{x'}-\mathbf{x})\ .
\end{equation}
From the study of the analytic structure of the longitudinal propagator
it is possible to calculate the longitudinal gap solving the following
equation for $\Delta_L$\cite{EMPR_dic00}
\begin{equation}
\Delta_{L}^{2}=\Delta_{T}^{2}+\frac{(m_s/S)^2}{\Gamma_{1}(0,\Delta_{L})}\ ,
\end{equation}
where $\Gamma_{1}(0,\Delta_{L})$ is the real part of the ``polarization bubble''
(in $q=0$ and $\omega=\Delta_{L}$) and may be written as
\begin{equation}
\Gamma_{1}(0,\Delta_{L})=\frac{3}{4}g\int\limits_{2\Delta_{T}}^{+\infty}
\frac{d\omega}{\pi}\frac{1}{\sqrt{(\omega^{2}-4\Delta_{T}^{2})}}
\frac{1}{\omega^{2}-\Delta_{L}^{2}}\ .
\end{equation}
For the field strength that we have estimated for CsNiCl$_3$, we find
the value $\Delta_L=0.779$, which is greater than $\Delta_T$, as we
expected. It is important to stress that the presence of the continuum makes
inapplicable the SMA, which establishes the relation
$\chi_L=Sgc/\Delta_L^2$ between the longitudinal susceptibility $\chi_{L}=dm_{s}/dH_{s}$
and the gap.

\begin{figure}[h]
\vspace{8mm}
\centerline{\includegraphics[width=7.5cm]{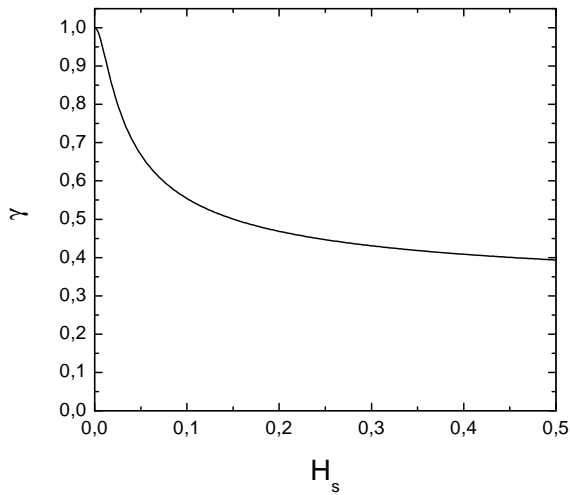}}
\caption{The relative quasi-particle weight of the pole of the longitudinal
propagator as a function of the staggered field, at $q=0$.}
\label{prefactor}
\vspace*{-10mm}
\end{figure}

For $\omega<2 \Delta_T$ the dynamical structure factor in the longitudinal channel
has well-defined poles corresponding to single particle excitations
\begin{equation}
S_L(q,\omega)=\gamma \frac{gcS^{2}}{2\varepsilon_L(q)}\left\{\delta(\omega-\varepsilon_L(q))-
\delta(\omega+\varepsilon_L(q))\right\}
\end{equation}
where the prefactor $\gamma$ is less than unity and gives
the reduction of the quasi-particle weight.
From the calculation\cite{EMPR_dic00}, it turns out that $\gamma$ is a
decreasing function of the staggered field, as displayed in Fig.\ref{prefactor}.

\begin{figure}[t]
\vspace{1mm}
\centerline{\includegraphics[width=7.5cm]{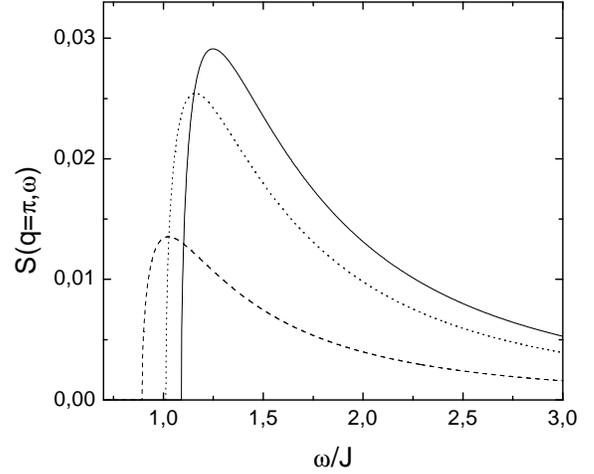}}
\caption{Continuum contribution to the spin dynamical structure factor
at the antiferromagnetic point, for $H_s=0.027$ (solid line), $H_s=0.020$ (dotted line) and
$H_s=0.010$ (dashed line). The integrated intensity grows as the staggered field increases.}
\label{s_pi_omega}
\vspace{5mm}
\end{figure}

As the field increases, the spectral weight that is lost from the pole
gets transferred to the multi-particle continuum, as required by the
first moment sum rule\cite{EMPR_dic00} of the dynamical structure factor.
We can observe this phenomenon by calculating the continuum contribution to
the dynamical structure factor for some values of the staggered field.
In Fig.\ref{s_pi_omega} the quantity $S(q, \omega)$ is plotted as a function
of $\omega$ at the antiferromagnetic point, for $H_s=0.027$, $H_s=0.020$
and $H_s=0.010$ (dashed line). The integrated intensity
grows as the staggered field increases.
We also notice that this contribution starts at the two-magnon
threshold $\omega>2 \Delta_T(H_s)$, consistently with experimental
data\cite{Sperim01,Ray} in which the continuum starts well before
the three magnon contribution as predicted by the NL$\sigma$M for a single chain\cite{Essler}.

At $H_s=0.027$ the value of the prefactor is $\gamma=0.746$, hence using the first moment
sum rule we find that the relative weight of the incoherent
contribution to the structure factor is
\begin{equation}
\frac{1-\gamma}{3}\approx 8.5\%
\end{equation}
which is a considerable amount, very close to the experimental measure of $9(2)\%$.

The present theory predicts the existence of a transversal mode and a longitudinal one with
different energies. This fact would be consistent with the presence of two well-defined
peaks in the inelastic scattering measurements with unpolarized neutrons.
However, in the experiments only one peak is observed.
Remembering that a similar problem arise in the case of $R_2$BaNiO$_5$\cite{Ray},
we argue that the longitudinal mode could be masked by its proximity to the
transversal one, by its reduced weight (less than 1/2 of the transverse weight) and
by the wide broadening.


In conclusion, we have proposed a theoretical mechanism to explicate
the anomalous multiparticle excitation weight observed in recent neutron
scattering experiments. This quantity turns out to be much greater than
the theoretical prediction on isolated $S=1$ chains.
We assume the existence of a temperature region above $T_N$, where the 3D effects
are not at all negligible, although the long range
order is destroyed, and we take into account of the interaction
with the neighboring chains by means of a staggered magnetic field.
We found that, even in the framework of a simple mean-field theory,
this scenario may account for the generation of a sizable continuum weight,
as well as an upward gap renormalization and a reduction of the correlation length,
breaking also the SMA correspondence between these two quantities.
Further and more detailed analytical calculation and numerical tests will be
treated in a future paper.

\subsection*{Acknowledgments}
We would like to thank A.A.~Nersesyan, P.~Pieri and G.~Sierra for useful discussions.

\vspace*{2cm}

\begin{center}
-----------------------------------------
\end{center}



\end{document}